*Full version for ArXiv. Shortened version to be submitted to PRL*

# Missed surface waves in non-piezoelectric solids

*Eugene A. Eliseev,*[1] *Anna N. Morozovska,*[2*] *Maya D. Glinchuk*[1], *and Sergei V. Kalinin*[3,†]

[1]Institute for Problems of Materials Science, National Academy of Sciences of Ukraine,
3, Krjijanovskogo, 03142 Kyiv, Ukraine

[2]Institute of Physics, National Academy of Sciences of Ukraine,
46, Prospekt Nauky, 03028 Kyiv, Ukraine

[3]Center for Nanophase Materials Science, Oak Ridge National Laboratory, Oak Ridge, Tennessee 37831, USA

## Abstract

The physical processes taking place at the surface and near the surface of solids is so rich and versatile that sometimes they seem to be the inexhaustible subject of fundamental research. In particular, since the discovery by Lord Rayleigh surface waves in solids focus increased attention of scientists, because their experimental and theoretical studies can serve as the source of unique information about the surface impact on the dynamics and structure of the atomic lattice, structural instabilities and phase transitions induced by the surface, and explore the properties of phonons in spatially-confined systems The existence of purely shear surface wave is impossible in non-piezoelectrics within the framework of the classical theory of elasticity, because the Rayleigh surface waves have different polarization and are the mixture of shear and dilatational waves.

We showed that the "forbidden" shear surface wave (shortly Flexo-SW) can propagate near the surface of all crystalline dielectrics due to the omnipresent flexoelectric coupling. The appearance and penetration depth of the Flexo-SW is ruled by the flexocoupling strength. In particular the penetration depth of transverse acoustic mode diverges in the absence of the flexoelectric coupling and so these waves become indistinguishable from the bulk waves. For paraelectrics such as strontium titanate with typical flexoeletric coefficients (~ 2V) the penetration depth of the Flexo-SW can reach more than tens microns at THz frequencies. The circumstances can explain the absence of experimental observations of the missed surface waves in thick layers and bulk materials. However we predict that the peaks of neutron intensity corresponding to the surface Flexo-SW and bulk phonon modes can be separated in non-piezoelectric thin films of thickness ~(20 – 50)nm.

* Corresponding author. E-mail: anna.n.morozovska@gmail.com (A.N.M.)

† Corresponding author. E-mail: sergei2@ornl.gov (S.V.K.)

## I. INTRODUCTION

**A. Amazing dynamics of the surfaces of solids.** The physical processes taking place in the vicinity and at the surface of solids is so rich and versatile that sometimes seems to be the inexhaustible subject of fundamental research [1, 2]. In particular, since the discovery of surface waves in solids [3], they focus increased attention of scientists [4, 5], because their experimental and theoretical studies can serve as the source of unique information about the surface impact on the dynamics and structure of the atomic lattice [6], structural instabilities and phase transitions induced by the surface [7], and explore the properties of phonons in spatially-confined systems [8, 9]. In addition to the fundamental aspect, the surface oscillations and waves are indispensable for applications in modern nanoacoustics [10] and nanoplasmonics [11].

**B. Surface waves: from the discovery to nowadays.** Existence of the surface acoustic wave (**SAW**) had been predicted theoretically in solids of arbitrary symmetry (including isotropic case) at the end of the 19th century by Lord Rayleigh [3]. The main features of the "classical wave solution" is **(i)** a specific elliptical polarization of the elastic displacement vector that rotation plane contains the surface normal and propagation vector. **(ii)** The shear wave with polarization perpendicular to the abovementioned plane cannot propagate along the surface as a surface waves. **(iii)** The longitudinal-transverse Rayleigh SAW is the mixture of shear and dilatation waves of expansions and compressions, in contrast to the acoustic waves propagating in the bulk of a solid matter, which have two transverse shear and one longitudinal dilatational modes [12]. Only at the end of the 60s of the 20th century Bleustein [13] and Gulyaev [14] had shown that the purely shear surface waves can propagate in some solids without inversion center (in particular at definite crystallographic cuts of piezoelectrics), and naturally their appearance is impossible in all non-piezoelectrics. Nonlinear Rayleigh waves that propagate along the surface of a homogeneous solid medium covered by a thin film had been considered in 1998 by Eckl et al [15]. The influence of a standing SAW on the diffusion of an adatom was theoretically studied in 2011 by Taillan et al [16]. Recently the interest to the theoretical consideration of classical linear SAW was renewed by Romero [17], who considered several types of SAW in piezoelectrics with specific "natural" boundary conditions imposed.

The next question naturally arises: how to investigate experimentally the SAW and how to verify existing theoretical predictions [13-17] and many others? Since the frequency of the soft mode related optic and acoustic phonons in piezoelectric and paraelectric materials lays within THz region and the wave vectors in the range (0.05 – 5)$nm^{-1}$, the phonon spectra can be extracted from the inelastic neutron scattering experiments [18, 19, 20, 21, 22, 23] by a conventional procedure. Hamilton et al. [24] performed the first experiment demonstrating that SAW in quartz can be probed by diffraction of cold neutrons. Much earlier Soffer et al proposed an optical imaging method for direct observation and study of SAWs [25] at the nonpolar Y-cut of piezoelectric $LiNbO_3$ (typical manifestation of Bleustein and Gulyaev waves). Pyrak-Nolte et al. [26] made the first direct observation of a new class of elastic interface waves propagating along the discontinuity of a synthetic fracture in aluminum. De Lima Jr et al. [27] present the experimental observation of Bloch oscillations, the Wannier-Stark ladder, and Landau-Zener tunneling of SAW in perturbed grating structures on a solid substrate (at that the vertical surface displacement has been measured by interferometry). Fine aspects of the SAW can be explored by Brillouin [11, 28] and Raman [29, 30] scattering, and Surface Enhanced Raman scattering based on the incomplete internal reflection [31]. Thus the experimental methods of SAW observation are well-evolved and enough precise to probe their finest properties and to verify the most sophisticated theoretical predictions.

**C. Expected role of the flexocoupling on the surface waves.** It should be noted that static flexoelectric coupling, i.e. the appearance of polarization due to the strain gradient, or the appearance of strain due to the polarization gradient [32, 33, 34], was not taken into account in all known theories of SAW [3, 12-17]. Moreover the notion about dynamic flexoelectric coupling [35, 36, 37] impact on phonon spectra was absent until recently [38, 39]. Nevertheless, an elastic wave of any kind is inevitably accompanied by a periodic gradient of mechanical strain and stress. This gradient is proportional to the wave vector of the oscillation and is obviously small for longer wavelengths. For a media of arbitrary symmetry (including an isotropic one) the wave of the strain gradient will cause a wave excitation of electric polarization (i.e., the local polarization, the mean value of which is zero) due to the flexoelectric effect. The latter, in turn, will affect the elastic stresses associated with the wave due to the inverse flexoelectric effect. Thus the flexocoupling should influence the properties of surface waves in all solids, since it essentially affects the bulk phonon spectra in different ferroelectrics and paraelectrics [38, 39], and the influence should be more pronounced for shorter wavelengths.

**D. Research motivation and impact.** Recently using Landau-Ginzburg-Devonshire (**LGD**) approach Morozovska et al. [38, 39] demonstrate the significant influence of the flexocoupling on the spatially modulated phase appearance, and on optic and acoustic phonon spectra in the bulk ferroelectric and paraelectric phases of ferroelectrics $PbTiO_3$, $Sn_2P_2(S,Se)_6$, and paraelectric $SrTiO_3$. Motivated by these results we used LGD approach for SAW description in paraelectrics and revealed that the surface shear waves similar to the waves of Bleustein [13] and Gulyaev [14] can exist in dielectrics of any symmetry (e.g. in paraelectric $SrTiO_3$) and for an arbitrary orientation of the surface due to the flexoelectric coupling. The wave is represented by the oscillations of coupled shear strain and electric polarization. Below we named this type of waves as **Flexo-SW**.

## II. STATEMENT OF THE PROBLEM ALLOWING FOR THE FLEXOCOUPLING

LGD expansion of bulk ($F_V$) and surface ($F_S$) parts of Helmholtz free energy $F$ on the polarization vector and strain tensor components $P_i$ and $u_{ij}$ have the form [38-39]:

$$F_V=\int_V d^3r\left(\begin{array}{l}\frac{a_{ij}}{2}P_iP_j+\frac{a_{ijkl}}{4}P_iP_jP_kP_l-P_iE_i+\frac{g_{ijkl}}{2}\left(\frac{\partial P_i}{\partial x_j}\frac{\partial P_k}{\partial x_l}\right)-q_{ijkl}u_{ij}P_kP_l\\ -\frac{f_{ijkl}}{2}\left(P_k\frac{\partial u_{ij}}{\partial x_l}-u_{ij}\frac{\partial P_k}{\partial x_l}\right)+\frac{c_{ijkl}}{2}u_{ij}u_{kl}+\frac{v_{ijklmn}}{2}\left(\frac{\partial u_{ij}}{\partial x_m}\frac{\partial u_{kl}}{\partial x_n}\right)\end{array}\right), \tag{1a}$$

$$F_S=\int_S d^2r\frac{a_{ij}^S}{2}P_i\,P_j\ . \tag{1b}$$

The tensor $a_{ij}$ is positively defined for linear dielectric, and explicitly depends on temperature $T$ for ferroelectrics and paraelectrics. In particular a Barrett-type [40] formula $a_{ij}=\alpha_{ij}^T\left(T_q\coth\left(T_q/T\right)-T_C\right)$ is valid for quantum paraelectrics. Note, that for proper ferroelectrics $T_C>T_q$, while $T_C<T_q$ for the quantum paraelectrics like $SrTiO_3$. $T_C$ is the Curie temperature, $T_q$ is a characteristic temperature, which is called the temperature of quantum vibrations sometimes. All other tensors included in the free energy (1) are supposed to be temperature independent. Tensor $a_{ijkl}$ should be positively defined for the functional stability in paraelectrics and ferroelectrics; it can be neglected for linear dielectrics. Tensors $g_{ijkl}$ and $v_{ijklmn}$, which determine the magnitude of the gradient energy, are also regarded positively defined. Coefficients $q_{ijkl}$ are the components of electrostriction tensor; $c_{ijkl}$ are the components of elastic stiffness tensor. Polarization is conjugated with electric field $E_i$ that can include external and depolarization contributions (if any exists). The flexoelectric energy is written in the form of Lifshitz invariant $\frac{f_{ijkl}}{2}\left(P_k\frac{\partial u_{ij}}{\partial x_l}-u_{ij}\frac{\partial P_k}{\partial x_l}\right)$, $f_{ijkl}$ is the flexocoupling tensor.

Lagrange function is

$$L=\int_t dt\left(F-K\right), \tag{2}$$

where the kinetic energy $K$ is given by expression

$$K=\int_V d^3r\left(\frac{\mu}{2}\left(\frac{\partial P_i}{\partial t}\right)^2+M_{ij}\frac{\partial P_i}{\partial t}\frac{\partial U_j}{\partial t}+\frac{\rho}{2}\left(\frac{\partial P_i}{\partial t}\right)^2\right), \tag{3}$$

which includes the dynamic flexoelectric coupling tensor $M_{ij}$. $U_i$ is the elastic displacement and ρ is the density of a material. The strain components are related with the displacement derivatives as $u_{ij}=\frac{1}{2}\left(\frac{\partial U_i}{\partial x_j}+\frac{\partial U_j}{\partial x_i}\right)$.

Dynamic equations of state have the form of Euler-Lagrange (**E-L**) equations:

$$\frac{\delta L}{\delta P_i}=0, \qquad \frac{\delta L}{\delta U_i}=0. \tag{4}$$

The boundary conditions at mechanically free surface can be obtained from the variation of the free energy (1) on polarization and strain:

$$\left.\left(g_{kjim}n_k\frac{\partial P_m}{\partial x_j}+a_{ij}^S P_j+\frac{f_{jkim}}{2}u_{jk}n_m\right)\right|_S=0. \qquad \left.\sigma_{ij}n_j\right|_S=0. \tag{5}$$

Here $n_k$ are components of the external normal to the surface, elastic stress tensor $\sigma_{ij}=-\delta F_V/\delta u_{ij}$ satisfies the mechanical equilibrium equation $\partial\sigma_{ij}/\partial x_j=0$. The most evident consequence of the flexocoupling is the inhomogeneous terms in the boundary conditions (5).

## III. ANALYTICAL SOLUTION FOR A MISSED SURFACE WAVE

### A. Explicit form of the Euler-Lagrange boundary problem for transverse surface waves

Further let us consider a phonon wave propagating along direction $x_1$ near the surface of semi-infinite solid with the surface plane $x_3=0$. It is relevant to consider the wave of polarization component $P_2(x_1,x_3,t)$ as the one that does not create any depolarization field; and so the transverse elastic wave with the displacement component $U_2(x_1,x_3,t)$ is not damped by the depolarizing effects influence. Sketch of the considered wave geometry is shown in **Fig. 1.**

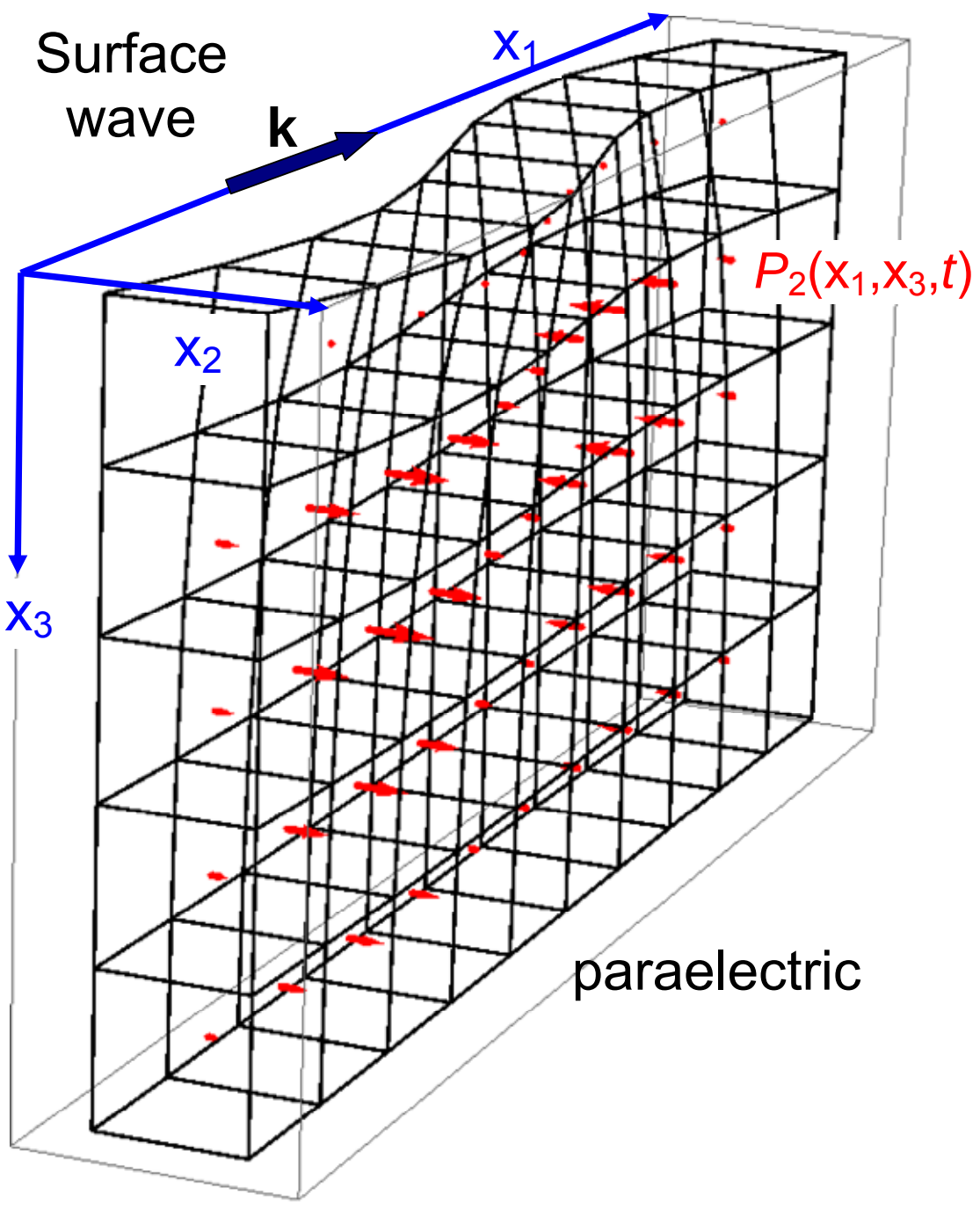

**FIG. 1.** Considered geometry of the surface wave propagation in a semi-infinite non-piezoelectric (dielectric or paraelectric) material. Red arrows are the elementary dipoles, which are zero at the surface in the particular case $\alpha_{S0}=\infty$ (because $P_2(x_1,0,t)=0$ at $\alpha_{S0}=\infty$). Black grid illustrates the deformation of the unit cells caused by the displacement $U_2(x_1,x_3,t)$ (the scale is distorted).

Explicit form of the E-L equations for the purely transverse surface waves with the boundary conditions at the surface $x_3=0$ are derived in **Appendix A** of the **Suppl. Mat.** For most of the cases one can neglect the polarization relaxation (i.e., to put $\Gamma=0$) and omit high order elastic strain gradient (i.e., to put $v_{ijklmn}=0$) if the flexoelectric coefficients are below the critical values $f^{cr}_{ijkl}$ [41, 42]. For the flexoelectric coefficients higher than the critical ones the spatially modulated phase occurs [39], at that the relationship $f^{cr}_{ijqs}f^{cr}_{klqs}\cong g_{ijmn}c_{klmn}$ is valid under the condition $v_{ijklmn}=0$ [39, 41, 42]. Hereinafter we regard that the dynamic flexoeffect tensor is diagonal, $M_{ij}=M\delta_{ij}$, and the inequality $M^2<\rho\mu$ should be valid for the time stability of kinetic energy [see Eq.(3)]. Below we use an isotropic approximation for the tensor coefficients $a_{ij}=\alpha\delta_{ij}$ and $a^S_{ij}=\alpha_{S0}\delta_{ij}$ ($\delta_{ij}$ is a Delta-Kroneker symbol).

Under these simplifications E-L equations (4) can be linearized in dielectrics and paraelectrics (i.e., at α>0). For considered geometry the linearized E-L equations along with the boundary conditions (5) acquire relatively simple form:

$$\rho\frac{\partial^2 U_2}{\partial t^2}+M\frac{\partial^2 P_2}{\partial t^2}-\Delta\left(c_{44}U_2+f_{44}P_2\right)=0 \qquad (6a)$$

$$\mu\frac{\partial^2 P_2}{\partial t^2}+M\frac{\partial^2 U_2}{\partial t^2}+\alpha P_2-\Delta\left(g_{44}P_2+f_{44}U_2\right)=0 \qquad (6b)$$

Boundary conditions (5) acquire the form

$$\left.\left(c_{44}\frac{\partial U_2}{\partial x_3}+f_{44}\frac{\partial P_2}{\partial x_3}\right)\right|_{x_3=0}=0, \qquad \left.\left(\alpha_{S0}P_2-g_{44}\frac{\partial P_2}{\partial x_3}-f_{44}\frac{\partial U_2}{\partial x_3}\right)\right|_{x_3=0}=0. \tag{6c}$$

**B. General expressions relating the amplitudes, frequency dispersion and penetration depth of the travelling surface waves**

Let us look for the solution of the linearized boundary problem (6) in the form of a travelling surface wave:

$$P_2(x_1,x_3,t)=\exp\left(i(kx_1-\omega t)-\xi x_3\right)\tilde{p}(k), \quad U_2(x_1,x_3,t)=\exp\left(i(kx_1-\omega t)-\xi x_3\right)\tilde{u}(k). \tag{7}$$

Here $k$ is the wave vector in the direction of the wave propagation, $\omega$ is its frequency, $\xi$ is the parameter, determining the inverse penetration depth of the wave. Since the solid occupies the semi-space $x_3 \geq 0$, only the exponents either vanishing or not increasing at $x_3 \to \infty$ are present, so $\mathrm{Re}(\xi)\geq 0$.

The substitution of expression (7) in Eqs.(6) leads to the system of linear algebraic equations for the amplitudes $\tilde{p}$ and $\tilde{u}$

$$\begin{cases}\left(\rho\omega^2+c_{44}\left(\xi^2-k^2\right)\right)\tilde{u}+M\,\omega^2\tilde{p}+f_{44}\left(\xi^2-k^2\right)\tilde{p}=0,\\ \left(\mu\,\omega^2-\alpha+g_{44}\left(\xi^2-k^2\right)\right)\tilde{p}+M\,\omega^2\tilde{u}+f_{44}\left(\xi^2-k^2\right)\tilde{u}=0.\end{cases} \tag{8}$$

Zero determinant of the system (8) gives the condition of the wave (7) existence

$$\left(\rho\omega^2+c_{44}\left(\xi^2-k^2\right)\right)\left(\mu\,\omega^2-\alpha+g_{44}\left(\xi^2-k^2\right)\right)=\left(M\,\omega^2+f_{44}\left(\xi^2-k^2\right)\right)^2. \tag{9}$$

The solution of Eq.(9) for the penetration depth of the wave is

$$\xi_{1,2}^2=k^2+\frac{-B(\omega)\pm\sqrt{B^2(\omega)-4\left(c_{44}g_{44}-\left(f_{44}\right)^2\right)C(\omega)}}{2\left(c_{44}g_{44}-\left(f_{44}\right)^2\right)}, \tag{10a}$$

where the functions $B(\omega)$ and $C(\omega)$ are given by expressions

$$B(\omega)=\left(\mu\,\omega^2-\alpha\right)c_{44}+\rho\omega^2 g_{44}-2M\,\omega^2 f_{44}, \tag{10b}$$

$$C(\omega)=\rho\omega^2\left(\mu\,\omega^2-\alpha\right)-\left(M\,\omega^2\right)^2. \tag{10c}$$

Substitution of the solution (7) rewritten in the explicit form $Q_2=\exp\left(i(kx_1-\omega t)\right)\left(q_1\exp(-\xi_1 x_3)+q_2\exp(-\xi_2 x_3)\right)$ (where the symbol $q=p$ for polarization $P$ or $q=u$ for the strain field $U$) to the equations (8) and to the boundary conditions (6c) leads to the two independent equations for the determination of penetration depths $\xi_i$:

$$\xi_1=\xi_2, \tag{11a}$$

$$\alpha_{S0}\left(\rho\omega^2-c_{44}k^2-c_{44}\xi_1\xi_2\right)=\left(\xi_1+\xi_2\right)\xi_1\xi_2\left(c_{44}g_{44}-\left(f_{44}\right)^2\right). \tag{11b}$$

Along with Eqs.(11) the following relation between the amplitude $p$ and $u$ should be valid:

$$u_i = -\left(\frac{M\,\omega^2 + f_{44}\left(\xi_i^2 - k^2\right)}{\rho\omega^2 + c_{44}\left(\xi_i^2 - k^2\right)}\right)p_i . \qquad (12)$$

If the shear strain wave is excited by polarization, its resonant enhancement at definite frequency ω is possible under the condition $\rho\omega^2(k) + c_{44}\left(\xi_i^2[k,\omega(k)] - k^2\right) = 0$. The dispersion law $\omega(k)$ will be derived and analyzed below.

The evident form of Eq.(11a) is equivalent to the condition of zero discriminant in Eq.(10a), namely

$$B^2(\omega) - 4\left(c_{44}g_{44} - (f_{44})^2\right)C(\omega) = 0 . \qquad (13)$$

Note, that the solution of Eq.(13) with respect to frequency is independent on the wave vector (similar solution was found by Romeo [17]). Romeo et al noted that for the case the "frequency dispersion" is limited to the discrete set of frequency values $\omega_n(k_n)$. However in the secular case $\xi_1 = \xi_2 = \xi$ the expressions (7) for the solution should be modified as $P_2 = (p_1 - p_2\xi x_3)\exp(i(kx_1 - \omega t) - \xi x_3)$ and $U_2 = (u_1 - u_2\xi x_3)\exp(i(kx_1 - \omega t) - \xi x_3)$ [see **Suppl. Mat.**]. A detailed consideration of the case presented in the part D of **Suppl. Materials** leads to the conclusion that the surface wave can exist under the validity of a very specific boundary condition, $\alpha_{S0} = 0$. Since it exists for a definite frequency only the "isolated" point unlikely can chances to be observed experimentally.

**C. Impact of the boundary conditions for polarization on the surface waves existence**

In contrast to the pessimistic scenario of the experimental verification of the secular case (11a), the solution of Eq.(11b) has sense at all $\alpha_{S0}$, and can be simplified for two limiting cases, considered below.

**(a)** The case of the natural boundary condition for polarization, when $\alpha_{S0} = 0$ and thus polarization normal derivative is zero at the surface $x_3 = 0$. Mathematically the case is equivalent to the condition $\xi_1\xi_2 = 0$, because $\xi_1 + \xi_2 \neq 0$. Setting $\xi = 0$ in Eq.(12) we immediately obtain the dispersion relation

$$\left(\rho\omega^2 - c_{44}k^2\right)\left(\mu\,\omega^2 - \alpha - g_{44}k^2\right) = \left(M\,\omega^2 - f_{44}k^2\right)^2 . \qquad (14)$$

In fact Eq. (14) represents dispersion relation for transverse phonon mode in the bulk, because the corresponding penetration depth given by Eqs.(10) is zero, $\xi = 0$. As anticipated Eq.(14) coincides with Eq.(13b) from Ref.[38] in a paraelectric phase with $P_S = 0$ and $2\alpha \rightarrow \alpha$ (due to the absence of factor ½ in the free energy in Ref.[38]).

**(b)** The case of zero polarization at the surface, when $\alpha_{S0} = \infty$ and thus $\rho\omega^2 - c_{44}k^2 - c_{44}\xi_1\xi_2 = 0$ from Eq.(12). The latter condition jointly with the condition $p_1 = -p_2$ is sufficient to satisfy the boundary conditions (6c). The explicit form of the condition is

$$\frac{\left(\rho\omega^2 - c_{44}k^2\right)^2}{c_{44}^2} = \frac{\left(\rho\omega^2 - c_{44}k^2\right)\left(\mu\,\omega^2 - \alpha - g_{44}k^2\right) - \left(M\,\omega^2 - f_{44}k^2\right)^2}{c_{44}g_{44} - \left(f_{44}\right)^2}. \qquad (15)$$

The most important is that under the absence of static ( $f_{44} = 0$ ) and dynamic ( $M = 0$ ) flexocoupling the dispersion equation (15) reduces to either the bulk dispersion $\rho\omega^2 = c_{44}k^2$ or to the separate frequency point $\frac{\rho\omega^2}{c_{44}} = \frac{\mu\,\omega^2 - \alpha}{g_{44}}$ meaning that "frequency dispersion" is limited to the discrete set of frequency values instead of the functional dependence $\omega(k)$.

In the presence of flexoelectric coupling the explicit form of Eq.(15) is a biquadratic equation

$$A(k)\omega^4 - Q(k)\omega^2 + \frac{\alpha c_{44}k^2}{c_{44}g_{44} - f_{44}^2} = 0, \qquad (16a)$$

where the functions $Q(k) = \frac{\alpha\rho}{c_{44}g_{44} - f_{44}^2} + \left(\frac{g_{44}\rho + c_{44}\mu - 2M\,f_{44}}{c_{44}g_{44} - f_{44}^2} - 2\frac{\rho}{c_{44}}\right)k^2$ and $A(k) = \frac{\rho\mu - M^2}{c_{44}g_{44} - f_{44}^2} - \frac{\rho^2}{c_{44}^2}$ are introduced. Since $c_{44}g_{44} > f_{44}^2$ for the system stability, and $\alpha > 0$ for dielectrics and paraelectrics, the last term in Eq.(16a) are positive for the these materials. Since we regard that $\rho\mu > M^2$ for the Lagrangian (2) stability, the first term $A(k)$ can be of arbitrary sign, but inevitably becomes positive under the condition $f_{44}^2 \to c_{44}g_{44}$, i.e., when the flexoelectric coefficient increases towards the critical value. So that Eq.(16a) has two roots - transverse optic (**TO**) and acoustic (**TA**) modes under the condition $Q(k) > 0$ and relatively high $f_{44}^2$. At $Q(k) < 0$ and $A(k) < 0$ it contains only one TA mode. Corresponding equations for the decay factors can be derived from Eqs.(10):

$$\xi^4 + R(k)\xi^2 + \frac{\left(\rho\omega^2 - c_{44}k^2\right)^2}{c_{44}^2} = 0, \qquad (16b)$$

where the function $R(k) = \frac{\left(c_{44}\mu + g_{44}\rho - 2M\,f_{44}\right)\omega^2 - \alpha(T)c_{44}}{\left(c_{44}g_{44} - f_{44}^2\right)} - 2k^2$ is introduced. Since the last term in Eq.(16b) is positive because of $c_{44}g_{44} > f_{44}^2$, the conditions for which both decay factors $\xi_i$ become real are $R(k) < 0$ and $c_{44}^2R^2(k) \geq 4\left(\rho\omega^2 - c_{44}k^2\right)^2$. Both $\xi_i$ are complex in the case $c_{44}^2 R^2(k) < 4(\rho\omega^2 - c_{44}k^2)^2$ , and purely imaginary in the case $c_{44}^2 R^2(k) \geq 4(\rho\omega^2 - c_{44}k^2)^2$ and $R(k) > 0$.

Expressions (13)-(16) are the formal analytical solution of the considered problem, but only Eq.(15) or its explicit form Eqs.(16) contain the "missed" surface wave. Below we explore the wave dispersion in transverse direction and its penetration from the surface using a concrete example of quantum paraelectric $SrTiO_3$.

## IV. FLEXOCOUPLING IMPACT ON SURFACE WAVES PROPERTIES IN NON-PIEZOELECTRIC SOLIDS

### A. Frequency dispersion, phase velocity and penetration depth of SAW in $SrTiO_3$

Using Eqs.(16) we calculated frequency dispersion $\omega(k)$ of the travelling wave vector for the case of paraelectric $SrTiO_3$ (**STO**) at temperatures (100 – 400) K. TA and TO modes penetrating in the bulk were calculated from Eq.(14). Numerical values of the STO material parameters have been extracted from the fitting of phonon spectra obtained from the neutron scattering [18] and dielectric permittivity data [43]. STO parameters are listed in **Table I.**

**Table I.** Numerical values of the **STO** material parameters

| **Description** | **Symbol and dimension** | **Numerical value for $SrTiO_3$** |
|---|---|---|
| Coefficient at $P^2$ | $\alpha(T)$ ($\times$C$^{-2}$·m J) | $\alpha_T\left(T_q \coth(T_q/T) - T_C\right)$ |
| Inverse Curie-Weiss constant | $\alpha_T$ ($\times 10^5$C$^{-2}$·m J/K) | 15 |
| Curie temperature | $T_C$ (K) | 30 |
| Characteristic temperature | $T_q$ (K) | 54 |
| Surface energy coefficient | $\alpha_{S0}$ ($\times$C$^{-2}$·J) | $\infty$ |
| LGD-coefficient at $P^4$ | $\beta$ ($\times 10^8$ JC$^{-4}$·m$^5$) | 81 |
| LGD-coefficient at $P^6$ | $\gamma$ ($\times 10^9$JC$^{-6}$·m$^9$) | 0 |
| Electrostriction coefficient | $q_{44}$ ($\times 10^9$J m/C$^2$) | 2.4 |
| Elastic stiffness coefficient | $c_{44}$ ($\times 10^{10}$ Pa) | 11 |
| Gradient coefficient at $(\nabla P)^2$ | $g_{44}$ ($\times 10^{-10}$C$^{-2}$m$^3$ J) | 0.5 (fitting parameter) |
| Elastic strain gradient $(\nabla u)^2$ | $v$ ($\times 10^{-9}$V s$^2$/m$^2$) | 0 (fitting parameter) |
| Static flexocoefficient | $f_{44}$ (V) | +2.1 (exp. value) |
| Dynamic flexocoefficient | $M$ ($\times 10^{-8}$Vs$^2$/m$^2$) | −1 (fitting parameter) |
| Kinetic coefficient | $\mu$ ($\times 10^{-18}$s$^2$m J) | 1.45 (fitting parameter) |
| Material density at norm. cond. | $\rho$ ($\times 10^3$ kg/m$^3$) | 4.930 at 120 K |
| Lattice constant | $a$ (nm) | $a_x=a_y=a_z=0.395$ at 120 K |

The dispersion curves of the lowest transverse acoustic (TA) and transverse optic (TO) surface phonon modes are shown by solid curves in **Fig. 2(a)**. TA and TO modes penetrating in the bulk are shown by dashed curves in **Fig. 2(a)**. The frequency of TO mode is rather high, and the minimal distance between the TO and TA modes is about 5 THz at $k \approx 1.2$ nm$^{-1}$. The modes interaction is very weak in STO that is typical for paraelectrics. For the sake of comparison the dispersion corresponding to the bulk TA and TO modes are presented by dashed curves in **Fig. 2(a)**. The difference between the dispersion curves for the bulk and surface TO modes is the most pronounced (~ (2 – 5)THz) for small $k < 0.5\,\text{nm}^{-1}$, and remained essential for all considered values of the wave vector $0 \le k \le 5\,\text{nm}^{-1}$. The difference between the frequency dispersion of the bulk and surface TA modes becomes noticeable only for the wave vector values $k > 2\,\text{nm}^{-1}$. Note that the differences between the bulk and surface TO modes decreases and the differences between the bulk and surface TA modes increases with the temperature increase [compare solid and dashed curves of different colors in **Fig. 2(a)**].

Despite the difference between the frequency dispersion of the surface and bulk TA modes are essentially smaller than between the corresponding TO modes, further we limit our consideration by surface TA modes properties, primary because their penetration depth is real [**Fig. 2(c)**] and the acoustic frequency is much lower that the optical one [**Fig. 2(a)**]. These properties of surface TA modes open the interesting possibilities for their excitation and experimental observation. In contrast, it appeared that TO modes penetration depth is purely imaginary for STO ($\xi = iq_z$) and so they are not localized near the surface. Namely the TO mode calculated from Eqs.(16) is a standing waves reflected from the surface $x_3 = 0$, and it disappears with the flexocoefficient $f_{44}$ decrease below 1.5 V. As a matter of fact the impact of the flexocoupling on the standing TO waves deserves a separate study, because their amplitude can be noticeable in thin films [see the next section].

The dependence of the surface TA wave penetration depths $1/\xi_1$ and $1/\xi_2$ on the wavelength $\lambda$ is shown in **Fig. 2(b)** for several temperatures (100 – 400) K. Because the penetration depth $1/\xi_1$ rapidly increases with the wavelength increase (see solid curves in **Fig. 2(b)**), the surface wave properties gradually tend to the ones of bulk wave in the limit $\lambda \to \infty$. The depth $1/\xi_2$ firstly increases, then reaches a very smooth maximum (or a plateau) and then saturates with the temperature decrease. Both penetration depths $1/\xi_1$ and $1/\xi_2$ almost coincide at small $\lambda < 1$ nm (compare dashed and solid curves in **Fig. 2(b)**). Note that the depth $1/\xi_1$ monotonically increases with the temperature increase; and the depth $1/\xi_2$ decreases with the temperature decrease (compare black, red, purple and blue dashed curves in **Fig. 2(b)**). Since the depths determine the localization of the surface wave, only the higher value $1/\xi_m = \max[1/\xi_1, 1/\xi_2]$ matters.

Dispersion law for Rayleigh, Bleustein and Gulyaev waves are similar for those of bulk elastic (infrasound/acoustic/ultrasound) waves, frequency of wave, $\omega$, is proportional to the wave number, $k$, namely $\omega = v_P\, k$, where $v_P$ is the wave velocity. The dependence of the surface TA wave phase velocity $v_P = \omega/k$ on its wavelength $\lambda = 2\pi/k$ is shown in **Fig. 2(c)** for several temperatures from the range (100 – 400) K. Firstly the phase velocity increases enough sharply and monotonically with the wavelength increase from 0.1nm to 10 nm, and then it saturates and tends to the phase velocity of the shear wave in the bulk of material. Smaller $v_P$ values correspond to the lower temperatures (compare black, red, purple and blue curves in **Fig. 2(c)**). The saturation starts at $\lambda$ values about 5 nm for $T = 400$ K, and about 30 nm for $T = 100$ K.

The frequency spectrum of the phase velocity is shown in **Fig. 2(d)** for several temperatures (100 – 400) K. The velocity monotonically decreases with the frequency increase at frequencies less than the critical value $\omega_{cr}$, at that $\omega_{cr} \approx 3.7$ THz at 100 K and $\omega_{cr} \approx 8.75$ THz at 400 K (compare black, red, purple and blue curves in **Fig. 2(d)**). At frequencies $\omega > \omega_{cr}$ the velocity is zero; hence the second-order phase transition occurs at $\omega = \omega_{cr}$. Numerical values of the phase velocity ~(1 – 4)km/s are in the same interval than the SAW velocity (3472.5 ± 1.5) m/sec measured by Soffer et al [25] at the nonpolar Y-cut of piezoelectric $LiNbO_3$ (typical manifestation of Bleustein and Gulyaev SAWs). However considered SAWs have the eigen frequencies $\omega(k) \sim 5$ THz at the wave vector $k = (1 - 100)$ nm, while the SAWs in $LiNbO_3$ were excited at resonant frequency about 40 MHz at $k = 0.1$mm. The several orders of magnitude difference questioned the possibility to observe and study SAWs in paraelectrics using the optical spatial filtering technique [25].

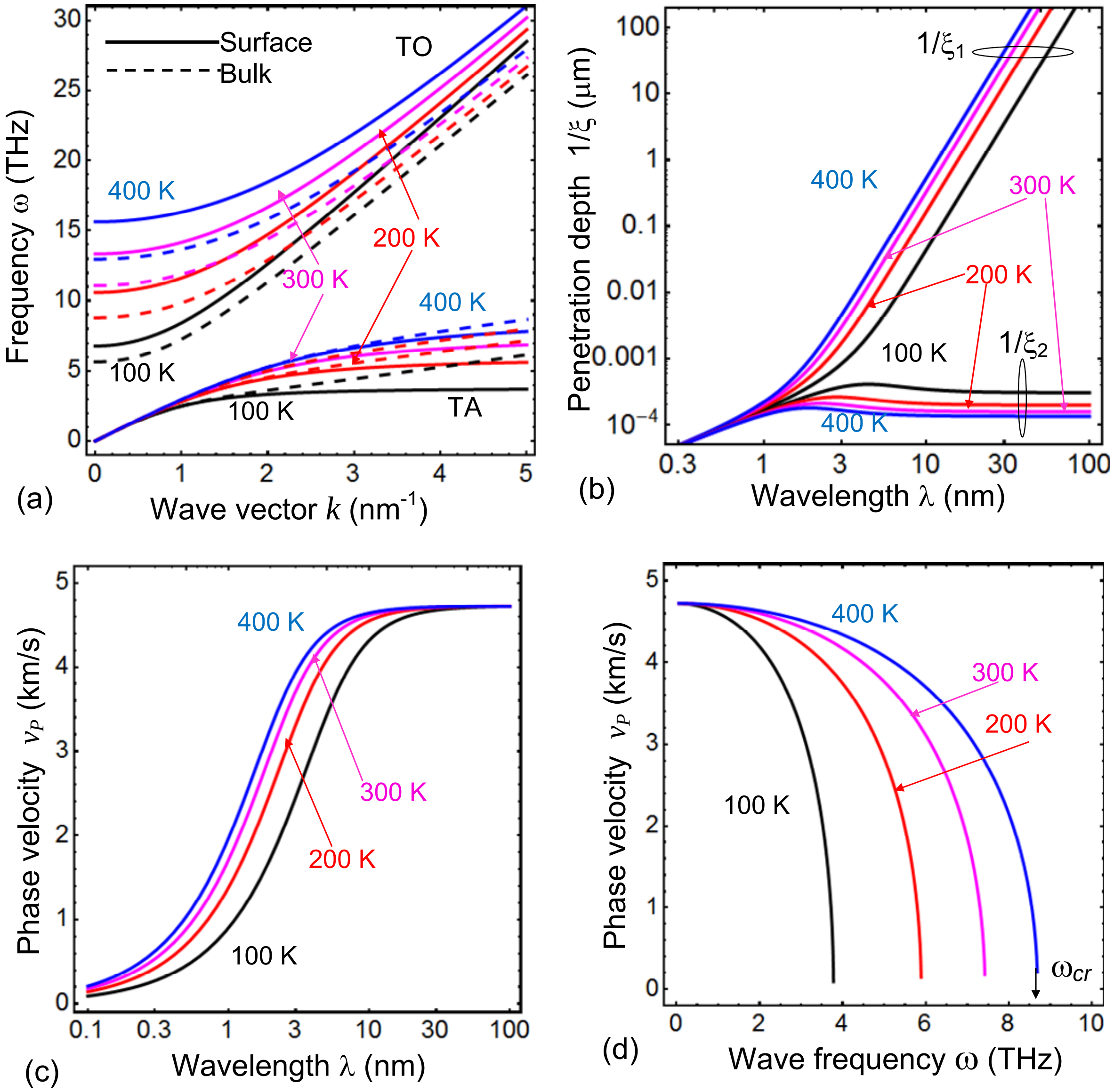

**FIG. 2. (a)** Frequency dispersion of the bulk (dashed curves) and surface (solid curves) phonon modes calculated for STO parameters. Transverse optic ("TO") and acoustic ("TA") modes are shown. **(b)** Dependence of penetration depth of the surface TA wave on its wavelength. **(c)** Phase velocity of the surface TA wave in dependence on the wavelength. **(d)** Phase velocity dependence on the surface wave frequency. Different curves in parts **(a)-(d)** correspond to the temperatures $T$=100, 200, 300 and 400 K, which values are indicated near the curves. Static flexoelectric coefficient $f_{44}$=2.1 V, dynamic flexocoupling constant $M=-1\times10^{-8}\mathrm{Vs^2/m^2}$, surface energy parameter $\alpha_{S0}=\infty$. Other material parameters of STO are listed in **Table I.**

## B. Impact of the flexocoupling on the SAW frequency dispersion and penetration depth

Note that **Fig. 2** is calculated for the static flexoelectric coefficient $f_{44}$=2.1 V and dynamic flexocoupling constant $M=-1\times10^{-8}\mathrm{Vs^2/m^2}$ extracted from the soft phonon spectra measured by neutron scattering by Yasusada and Shirane [18] [see **Fig. 5(a)**]. Note that the extracted value $f_{44}$=2.1 V is in a surprising agreement with the value $f_{44}$=(2.18±0.05) V determined from the bending of STO crystal by Zubko et al [44], but earlier they measured that $f_{44}$=1.3 V [45]. The dynamic flexocoupling constant absolute value $1\times10^{-8}$Vs$^2$/m$^2$ is inside the range $(0 - 20)\times10^{-8}$Vs$^2$/m$^2$ reported in Refs. [38, 39]. Since exact values of $f_{ij}$ and $M$ are still under debate for most of ferroics including ferroelectrics and quantum paraelectrics [46, 47, 48], it seems reasonable to explore the properties of the revealed surface TA wave on the value of $f_{ij}$ varying in the actual range (0 – 3)V. Hereinafter we consider $M<0$ for STO, because the inequality $M\,f_{44}<0$ is in a much better agreement with the phonon spectra [19, 20] and bending measurements [44] than the case $M\,f_{44}>0$. Results are presented in **Figs 3** and **4**. The case $M$=0 is shown in the figures for comparison.

**Figures 3(a)** and **3(b)** demonstrate the dependences of the penetration depth of the surface TA mode $1/\xi_m$ on the flexoelectric coefficient $f_{44}$ and wave vector $k$ calculated for the cases $M=0$ and $M\neq0$, respectively. **Figures 3(c)** and **3(d),** which are cross-sections of the **Figs. 3(a)** and **3(b)** at several fixed $k$**,** show the dependence of the $1/\xi_m$ on the $f_{44}$. Under the condition $M=0$, the penetration depth $1/\xi_m$ sharply increases (up to cm) with the flexoelectric coefficient $f_{44}$ decrease below 0.5 V and diverges when its value tends to zero [see different curves in **Fig. 3(c)** and contour lines in **Fig. 3(a)**]. When the penetration depth $1/\xi_m$ diverges, the surface wave properties coincide with the ones of a bulk wave. In contrast, for $f_{44}>1$V and $f_{44}<f_{44}^{cr}$, and wave vectors $k>1$ nm$^{-1}$ the depth $1/\xi_m$ becomes less than 100 nm, so it is indeed becomes a surface acoustic wave. The depth $1/\xi_m$ very sharply increases (up to the infinity) in the immediate vicinity $f_{44}\to f_{44}^{cr}$, and becomes imaginary at $f_{44}>f_{44}^{cr}$ indicating the onset of the spatially modulated phase. The critical value of the spatially modulated phase appearance is $f_{44}^{cr}=\sqrt{g_{44}c_{44}}\approx2.35$ V, and it is independent on the dynamic flexocoupling value as anticipated [39].

The divergence of $1/\xi_m$ at $f_{44}=0$ disappears for negative $M$ and positive $f_{44}$. Corresponding curves have a sharp maximum only at $f_{44}^{cr}$ [see different curves in **Fig. 3(d)** and contour lines in **Fig. 3(b)**]. Actually we established that the divergence $1/\xi_m$ can originate from the last term $\left(M\,\omega^2-f_{44}k^2\right)^2$ in Eq.(15) for the TA mode frequency ω. Since the term is positive for the case $f_{44}M<0$, corresponding penetration depths given by Eqs.(10) are finite. Negative sign of $f_{44}$ inducing to the additional divergence of $1/\xi_m$ at negative $M$ value satisfying the equality $M=f_{44}$ seem in a contradiction with the values extracted from the neutron scattering [19, 20] and bending [44] experiments in STO. However the condition $M\geq f_{44}$ and is far not excluded for other materials.

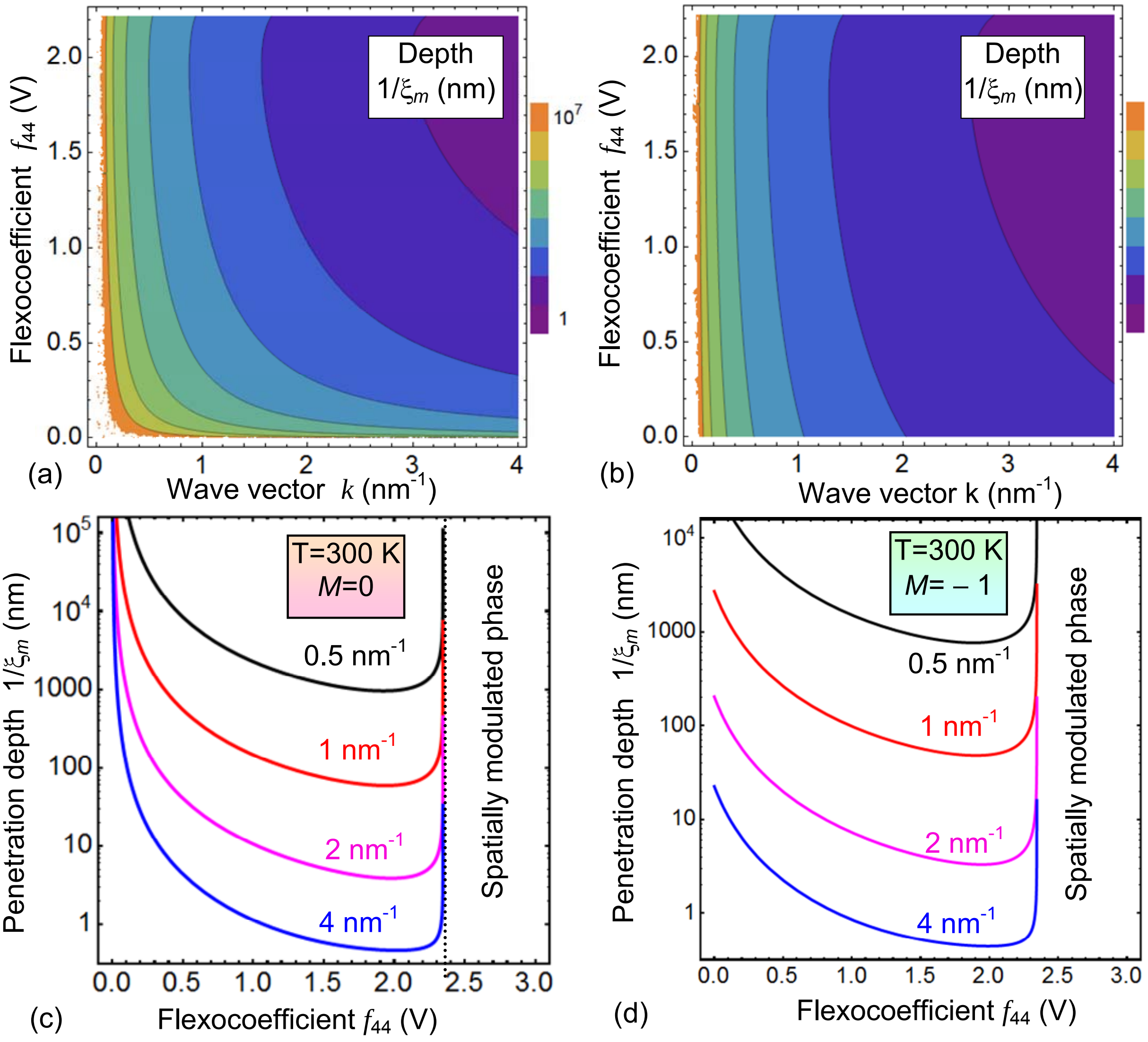

**FIG. 3.** Contour map of the TA mode penetration depth in coordinates "wave vector $k$ – flexoelectric coefficient $f_{44}$" calculated temperature for $M = 0$ **(a)** and $M = -4.8\times10^{-8}\text{Vs}^2/\text{m}^2$ **(b)**. Dependence of the TA wave penetration depth $1/\xi_m$ on the flexoelectric coefficient $f_{44}$ calculated at 300 K for $M = 0$ **(c)** and $M = -1\times10^{-8}\text{Vs}^2/\text{m}^2$ **(d).** Different curves in parts **(c)-(d)** correspond to the different values of the wave vectors $k$=0.5, 1, 2, 4 nm$^{-1}$, which values are listed near the curves. Other parameters corresponding to STO are listed in **Table I.**

**Figures 4** show the dependence of the surface TA mode frequency ω on the flexoelectric coefficient $f_{44}$ and wave vector $k$ calculated for the cases $M = 0$ [**Figs. 4(a)** and **4(c)**] and $M \neq 0$ [**Figs. 4(b)** and **4(d)**]. The difference between these cases are relatively small, leading to the conclusion that the impact of $M$ value on the frequency ω of surface TA mode is relatively small (at least in comparison with its influence on the penetration depth). For both M=0 and M<0 the frequency ω becomes higher than 1 THz for small wave vectors $k < 0.2$ nm$^{-1}$ and static flexocoefficients $f_{44}$ lying in the physically reasonable range (0 – 3) V. The frequency ω is almost independent on the flexoelectric coefficient $f_{44}$ for wave vectors $k < 1$ nm$^{-1}$ [see almost vertical contour lines of constant ω in **Figs. 4(a)** and **4(b)**]. The frequency values are relatively high (>4 THz) at $k > 0.5$ nm$^{-1}$, but such values are typical for the soft phonons frequency. Under the condition $k > 1$ nm$^{-1}$ the frequency relatively slowly and monotonically decreases with $f_{44}$ increase [**Figs. 4(c)** and **4(d)**]. Since the product $f_{44}M$ appeared negative for STO, no maxima are present at the dependence $\omega(f_{44})$ at fixed $k$ [**Figs. 4(d)**].

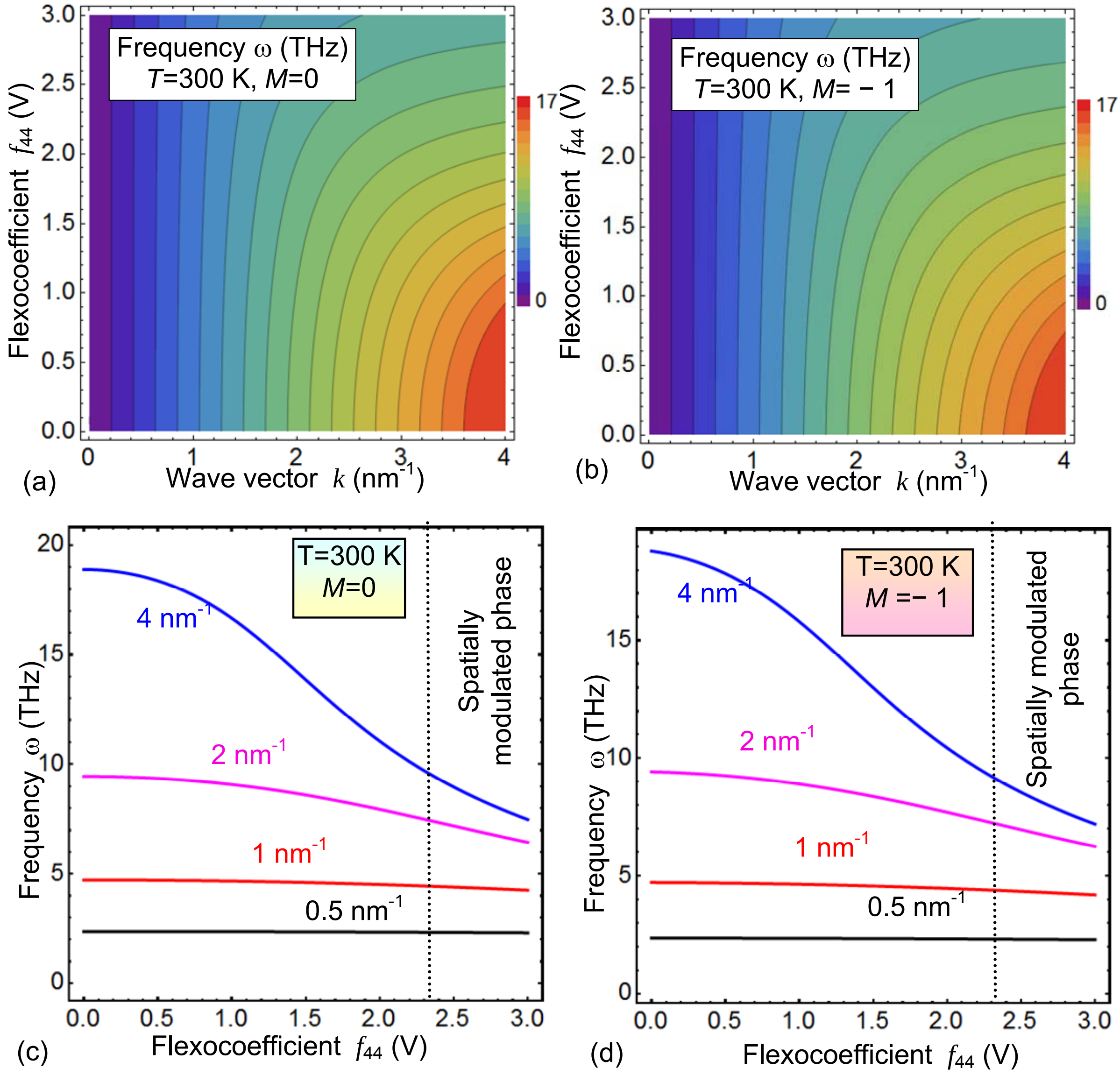

**FIG. 4.** Contour map of the TA mode frequency in coordinates "wave vector $k$ – flexoelectric coefficient $f_{44}$" calculated temperature for $M = 0$ **(a)** and $M = -1\times10^{-8}$Vs$^2$/m$^2$ **(b)**. Dependence of the surface TA wave frequency on the flexoelectric coefficient $f_{44}$ calculated at 300 K for $M = 0$ **(c)** and $M = -1\times10^{-8}$Vs$^2$/m$^2$ **(d).** Different curves in **(c)-(d)** correspond to the different values of the wave vectors $k$=0.5, 1, 2, 4 nm$^{-1}$, which values are listed near the curves. Other parameters corresponding to STO are listed in **Table I.**

To resume the analyses of the graphical results presented in section IV we can state that the existence and penetration depth of the revealed surface TA phonon mode is ruled by the static and dynamic flexocouplings. In particular the mode transforms to the bulk wave in the absence of the couplings. So the flexoelectricity indeed generates previously unexplored type of acoustic waves, further abbreviated as **Flexo-SW**, which can travel near the surface of any solid. Next we can speculate whether these surface waves be excited and detected separately from the classical bulk phonon modes.

## VI. POSSIBILITIES OF FLEXO-SW EXCITATION AND EXPERIMENTAL OBSERVATION

Since the calculated frequency dispersion $\omega(k)$ of the Flexo-SW is within THz region for the wave vectors in the range $k$=(0.05 – 5)nm$^{-1}$ in non-piezoelectric paraelectrics with relatively small coefficient $\alpha \cong 1/\varepsilon_0\varepsilon$ corresponding to the high relative dielectric permittivity $\varepsilon \geq 100$, the waves can be excited similarly to the bulk acoustic phonons, and the dispersion $\omega(k)$ can be determined from inelastic neutron scattering [18, 19]. For instance the dispersion curves of the bulk and surface TO and TA modes in STO are shown in **Fig. 5(a)** for actual range of neutron energy ($5\text{meV} \leq 2\pi\hbar^2 k^2/m_n \leq 50\text{meV}$) and three different temperatures. We expect that the peaks of neutron intensity corresponding to the surface and bulk phonon modes can be separated in thin non-piezoelectric paraelectric layers. This is possible because the difference between the energy of the surface and bulk phonons is $\Delta E_{TO}(k) = (1 - 3)$ meV for TO modes at $k$=(0.1 – 5)nm$^{-1}$, and $\Delta E_{TA}(k) = -(0.5 - 3)$ meV for TA modes at $k$>2nm$^{-1}$ at temperatures (100 – 300) K [see **Fig. 5(b)**]. Corresponding penetration depth of TA mode $\xi^{-1}(k)$ becomes is about or less than 10 nm at $k$>1 nm$^{-1}$ and $T$=100 K, and at $k$>2 nm$^{-1}$ and T=300 K [see **Fig. 5(c)**], and so we expect that the surface and bulk phonon modes can be separated in thin non-piezoelectric layers with thickness of several penetration depths $\xi$ (for both surfaces to contribute into the response), which is about or less than 50 nm for STO.

TO modes, which "penetration depth" appeared purely imaginary for STO parameters [see **Fig. 5(d)**], can be imagined as standing TO waves reflected from the surface $x_3 = 0$. As it was mentioned, TO mode disappears with the flexocoefficient $f_{44}$ decrease below 1.5 V. The standing TO waves deserves to be noticeable in thin films, which thickness is an integral multiple of their period $2\pi/q_z(k)$.

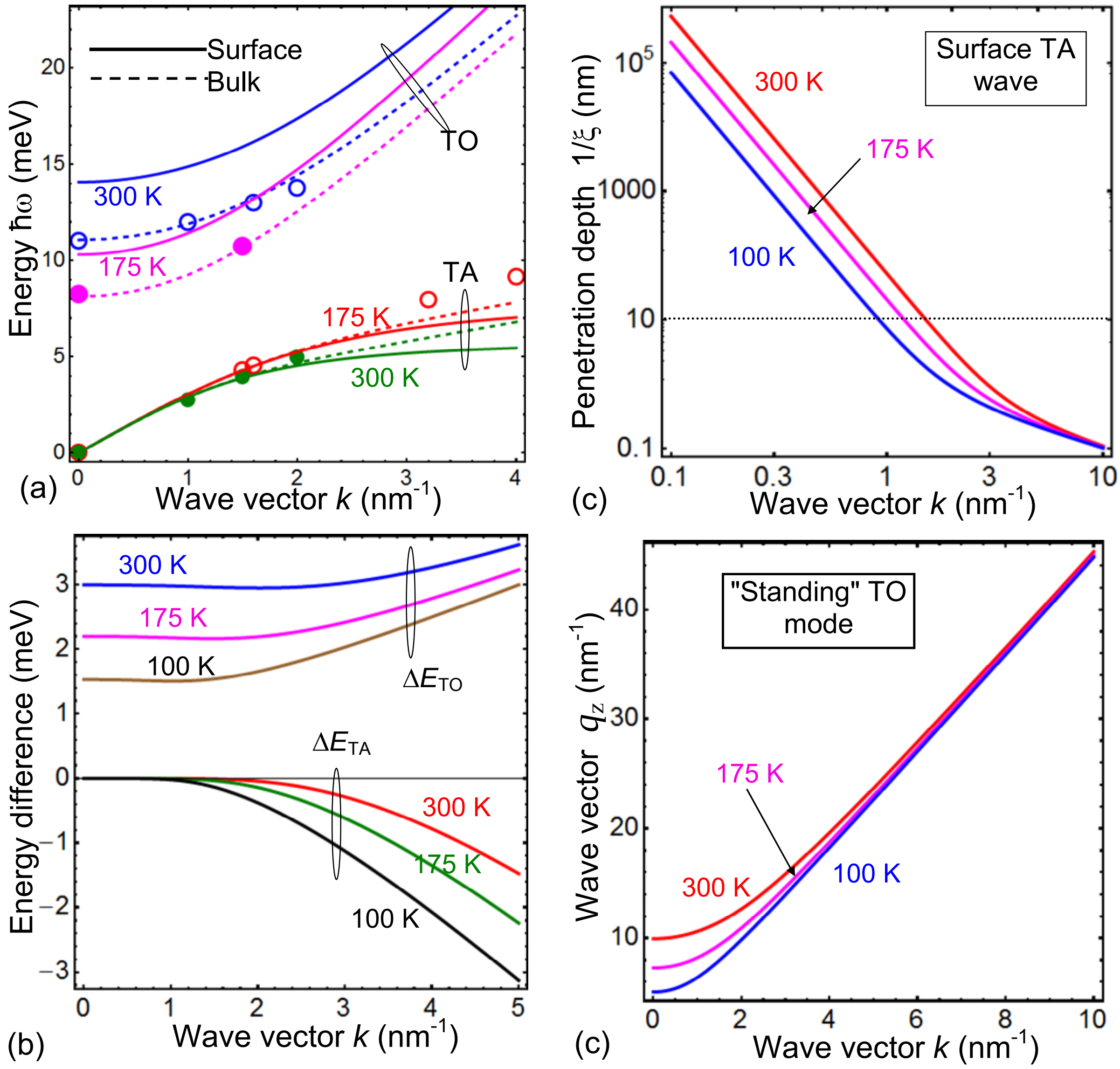

**FIG. 5. (a)** Energy dispersion $E(k)$ of the bulk (dashed curves) and surface (solid curves) phonon modes calculated in STO (solid curves). Symbols are experimental data [201]. **(b)** Energy difference $\Delta E(k)$ of the surface and bulk TO modes (top curves with label $\Delta E_{TO}$) and TA modes (bottom curves with label $\Delta E_{TA}$). **(c)**. Dispersion of the surface TA mode penetration depth $\xi^{-1}(k)$ calculated in STO. **(d)** Dispersion of the TO modes wave vector $q_z(k)$ in the direction $x_3$, normal to the surface. Corresponding localization depth is purely imaginary, $\xi = iq_z$. Different curves in parts **(a)-(d)** correspond to the temperatures $T$=100, 200, 300 and 400 K, which values are indicated near the curves. STO parameters obtained from the fitting of experimental data [20] are listed in **Table I.**

The dispersion of the TA shear strain wave amplitudes $u_i(k)$ calculated from Eq.(12) are shown in **Fig.6(a)**. The amplitude is normalized on the polarization amplitudes $p_i$ regarded proportional to the applied electric field $\mathbf{E}_0$, $p_i \sim \chi_{ij} E_j^0$. Contour maps of the TA wave amplitudes of polarization $P_2(x_1,x_3,t)$ and displacement $U_2(x_1,x_3,t)$ components are shown in **Figs.6(b)** and **6(c)**, respectively. The maps were calculated from Eq.(7) for fixed frequency $\omega(k)$, time moment $t$ and wave vector $k = 1$ nm$^{-1}$. **Fig. 6(d)** shows the atomic displacements of Sr and Ti near the surface of $SrTiO_3$. As one can see from **Figs.6** the polarization wave is zero at the surface $x_3 = 0$ for the case $\alpha_{S0} = \infty$, because $p_1 = -p_2$. The wave amplitude has a maximum at depth $x_3 \approx 2$ nm and becomes negligibly small at $x_3 > 15$ nm. Displacement is maximal at $x_3 = 0$ and exponentially vanishes at $x_3 > 5$ nm. So that the neutron scattering in thin STO films of thickness less than (20 – 50)nm should give us the information about the surface TA phonons coupled with flexoelectricity.

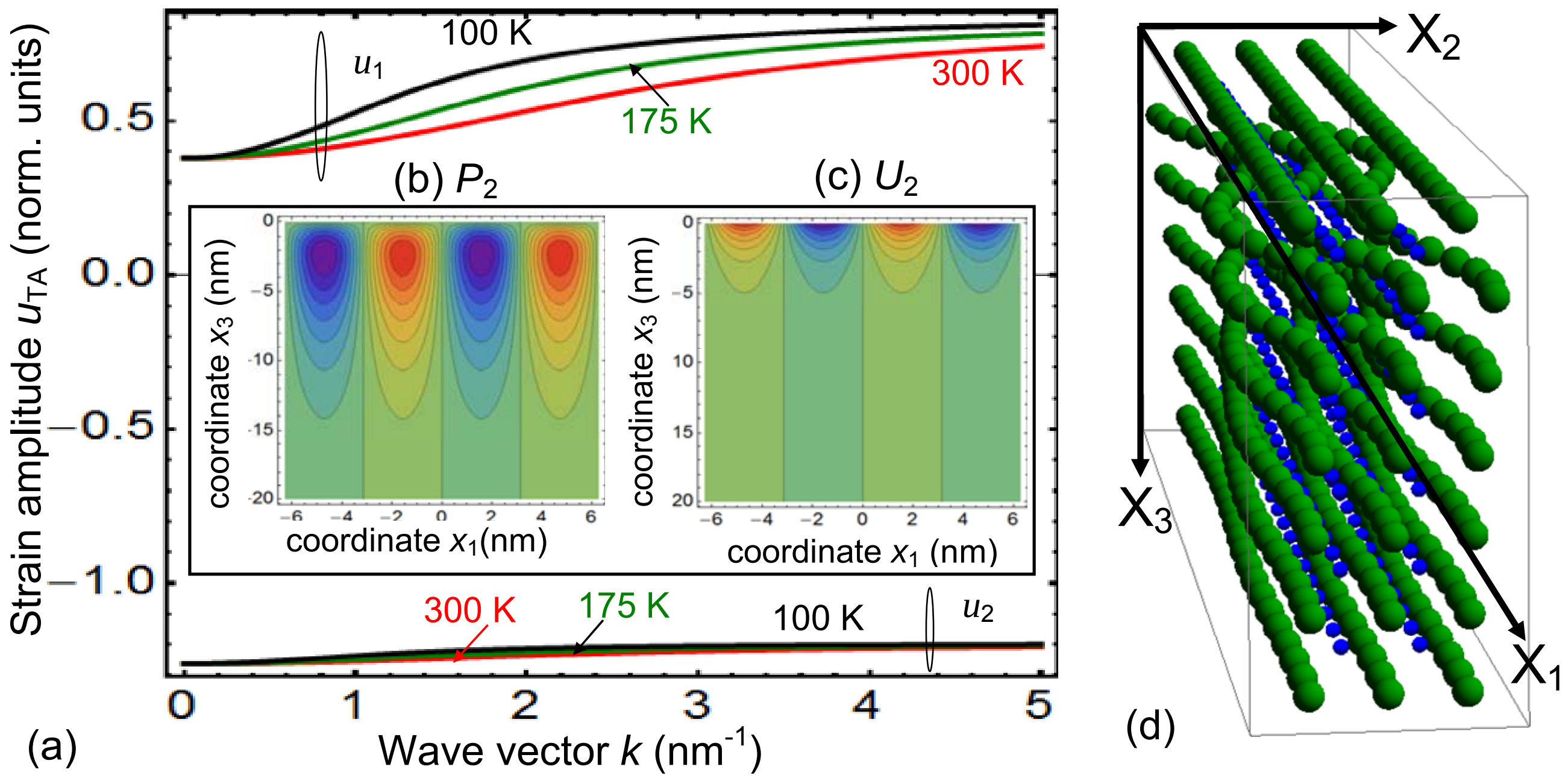

**FIG. 6. (a)** The dispersion of the normalized TA wave strain amplitudes $u_i(k)$ calculated in STO at temperatures 100, 175 and 300 K, which values ate shown near the curves. The amplitude is normalized on the polarization amplitudes $p_i$ . Insets: contour maps of the TA wave amplitudes $P_2(x_1,x_3,t)$ **(b)** and $U_2(x_1,x_3,t)$ **(c)** calculated from Eq.(7) for fixed frequency $\omega(k)$, time moment $t$ and wave vector $k = 1$ nm$^{-1}$. STO parameters are listed in **Table I. (d)** Atomic displacements of Sr (green balls) and Ti (blue balls) near the surface of $SrTiO_3$.

To resume the section, the possibilities of the Flexo-SW observation by inelastic neutron scattering is much more favorable in thin layers of paraelectrics with dielectric permittivity ε>>100 (i.e. in STO or KTO at low temperatures) in comparison with linear low-k dielectrics with ε≤10. Also we hope that some of the predicted properties of Flexo-SW can be verified and explored using optical imaging, infrared spectroscopy, Raman and Brillouin scattering and Surface Enhanced Raman scattering based on the incomplete internal reflection.

## VII. CONCLUSION

We showed that the shear surface wave (shortly Flexo-SW) can propagate near the surface of all crystalline dielectrics with the flexoelectric coupling. The existence of such a wave is not possible in non-piezoelectrics within the framework of the classical theory of elasticity, because Rayleigh waves have different polarization and are the mixture of shear and dilatational waves. Appeared that the penetration depth of the Flexo-SW depends essentially on the flexoelectric coefficients. In particular the penetration depth of acoustic mode diverges in the absence of the flexoelectric coupling and in the case these waves become indistinguishable from the bulk waves. Furthermore, it should be noted that for a paraelectric strontium titanate the localization depth of Flexo-SW can reach more than tens microns for the transverse wavelength of more than tens nm and THz frequencies. The circumstances can explain the absence (up to now) of experimental observations of the predicted surface waves. We expect that the peaks of neutron intensity corresponding to the surface acoustic Flexo-SW and bulk phonon modes can be separated in thin non-piezoelectric paraelectric layers, giving a unique opportunity to define the flexoelectric coefficients.

We also obtained that revealed acoustic Flexo-SW has a rather complicated dispersion law that main features are following:

1. With decreasing the wave vector $k$ the wave velocity along the surface approaches the speed of bulk shear waves, while the penetration depth tends to infinity.
2. For wavelengths about micrometer order and less the phase velocity of the surface wave decreases, and its penetration depth increases up to tens of microns.
3. The dispersion relation for Flexo-SW depends strongly on the boundary conditions for the electric polarization at the surface of the material, so that more detailed investigation is required.

In contrast, it appeared that the penetration depth of transverse optic modes is purely imaginary for the strontium titanate and so they are not localized near the surface. In fact the mode is a standing waves reflected from the surface $x_3 = 0$, and it disappears with the flexocoefficient decrease. The impact of the flexocoupling on the standing waves deserves a separate theoretical study, because their amplitude can be noticeable in thin paraelectric films.

**Authors' contribution.** E.A.E. and A.N.M. contributed equally to the research idea, problem statement and analytical calculations; E.A.E. prepared graphics and A.N.M. wrote the manuscript draft. M.D.G. worked densely on the results analyses and discussion. S.V.K. contributed very significantly to the research motivation, results analyses and discussion, manuscript text and structure improvement.

**Notice:** This manuscript has been authored by UT-Battelle, LLC, under Contract No. DE-AC0500OR22725 with the U.S. Department of Energy. The United States Government retains and the publisher, by accepting the article for publication, acknowledges that the United States Government retains a non-exclusive, paid-up, irrevocable, world-wide license to publish or reproduce the published form of this manuscript, or allow others to do so, for the United States Government purposes. The Department of Energy will provide public access to these results of federally sponsored research in accordance with the DOE Public Access Plan (http://energy.gov/downloads/doe-public-access-plan).

**Link to the Supplement to the manuscript [49]**

## SUPPLEMENT

## APPENDIX A. Derivation of dispersion relations

### A. Free energy and boundary conditions

Let us consider a purely transversal surface acoustic wave, which is impossible in "linear" elastic media without piezoelectric effect. Free energy functional relevant to the considered system is

$$F = F_V + F_S \tag{A.1a}$$

It consists of bulk, $F_V$, and surface, $F_S$, contributions, having the following form:

$$F_V = \int\limits_{x_3>0} dx_1 dx_3 \left( \begin{array}{l} \frac{\alpha(T)}{2}P_2^2 + \frac{\beta}{4}P_2^4 + \frac{\gamma}{6}P_2^6 - P_2 E + \frac{g_{1212}}{2}\left\{ \left(\frac{\partial P_2}{\partial x_1}\right)^2 + \left(\frac{\partial P_2}{\partial x_3}\right)^2 \right\} \\ -\frac{f_{1212}}{2} 2\left( P_2 \frac{\partial u_{12}}{\partial x_1} - u_{12}\frac{\partial P_2}{\partial x_1} + P_2 \frac{\partial u_{23}}{\partial x_3} - u_{23}\frac{\partial P_2}{\partial x_3} \right) \\ +\frac{c_{1212}}{2} 4\left(u_{12}^2 + u_{23}^2\right) + \frac{v_{112211}}{2} 4\left( \left(\frac{\partial u_{12}}{\partial x_1}\right)^2 + \left(\frac{\partial u_{23}}{\partial x_3}\right)^2 \right) + v_{112233}\frac{\partial u_{12}}{\partial x_1}\frac{\partial u_{23}}{\partial x_3} \end{array} \right) \tag{A.1b}$$

$$F_S = \int dx_1 \frac{\alpha_S}{2} P_2^2 \Big|_{x_3=0} \tag{A.1c}$$

Factors "2" and "4" takes into account that summation is performed with symmetric tensor $u_{ij} = u_{ji}$. We suppose that only one component of displacement matters, which depends on two coordinates, i.e. $U_2(x_1, x_3)$. Therefore, only two components of the strain tensor differ from zero, namely:

$$u_{12} = \frac{1}{2}\frac{\partial U_2}{\partial x_1}, \qquad u_{23} = \frac{1}{2}\frac{\partial U_2}{\partial x_3}. \tag{A.2}$$

Introducing Voight notations for tensors of material constants and using relations (A.2), we could rewrite Eq.(A.1b) as follows

$$F_V = \int\limits_{x_3 \geq 0} dx_1 dx_3 \left( \begin{array}{l} \frac{\alpha(T)}{2}P_2^2 + \frac{\beta}{4}P_2^4 + \frac{\gamma}{6}P_2^6 - P_2 E + \frac{g_{44}}{2}\left\{ \left(\frac{\partial P_2}{\partial x_1}\right)^2 + \left(\frac{\partial P_2}{\partial x_3}\right)^2 \right\} \\ -\frac{f_{44}}{2}\left( P_2 \frac{\partial^2 U_2}{\partial x_1 \partial x_1} - \frac{\partial U_2}{\partial x_1}\frac{\partial P_2}{\partial x_1} + P_2 \frac{\partial^2 U_2}{\partial x_3 \partial x_3} - \frac{\partial U_2}{\partial x_3}\frac{\partial P_2}{\partial x_3} \right) \\ +\frac{c_{44}}{2}\left( \left(\frac{\partial U_2}{\partial x_1}\right)^2 + \left(\frac{\partial U_2}{\partial x_3}\right)^2 \right) + \frac{v_{112211}}{2}\left( \left(\frac{\partial^2 U_2}{\partial x_1^2}\right)^2 + \left(\frac{\partial^2 U_2}{\partial x_3^2}\right)^2 \right) \\ +v_{112233}\frac{\partial^2 U_2}{\partial x_1^2}\frac{\partial^2 U_2}{\partial x_3^2} \end{array} \right) \tag{A.3}$$

Equations of state obtained from the variation of the free energy (A.3) on polarization $P_2$ and strain components $u_{12}$ and $u_{23}$ have the form:

$$\mu\frac{\partial^2 P_2}{\partial t^2} + M\frac{\partial^2 U_2}{\partial t^2} + \alpha(T)P_2 + \beta P_2^3 + \gamma P_2^5 - g_{44}\Delta P_2 - f_{44}\left( \frac{\partial^2 U_2}{\partial x_1 \partial x_1} + \frac{\partial^2 U_2}{\partial x_3 \partial x_3} \right) = 0, \tag{A.4a}$$

$$\sigma_{12} = c_{44}\frac{\partial U_2}{\partial x_1} + f_{44}\frac{\partial P_2}{\partial x_1} - v_{121}\frac{\partial^3 U_2}{\partial x_1^3} - v_{123}\frac{\partial^3 U_2}{\partial x_1 \partial x_3^2}, \tag{A.4b}$$

$$\sigma_{23} = c_{44}\frac{\partial U_2}{\partial x_3} + f_{44}\frac{\partial P_2}{\partial x_3} - v_{121}\frac{\partial^3 U_2}{\partial x_3^3}. - v_{123}\frac{\partial^3 U_2}{\partial x_3 \partial x_1^2}. \tag{A.4c}$$

Dynamical equations for elastic subsystem are obtained from the variation of Lagrange function (2):

$$\rho\frac{\partial^2 U_2}{\partial t^2}+M\frac{\partial^2 P_2}{\partial t^2}-c_{44}\Delta U_2-f_{44}\Delta P_2+v_{121}\left(\frac{\partial^4 U_2}{\partial x_1^4}+\frac{\partial^4 U_2}{\partial x_3^4}\right)+v_{123}2\frac{\partial^4 U_2}{\partial x_1^2\partial x_3^2}=0. \quad \text{(A.5)}$$

For the sake of simplicity below we neglect higher order gradients of displacement ( $v_{ijk}=0$ ). Therefore, Eqs.(A.4) and (A.5) could rewritten as

$$\rho\frac{\partial^2 U_2}{\partial t^2}+M\frac{\partial^2 P_2}{\partial t^2}-c_{44}\Delta U_2-f_{44}\Delta P_2=0, \quad \text{(A.6a)}$$

$$\mu\frac{\partial^2 P_2}{\partial t^2}+M\frac{\partial^2 U_2}{\partial t^2}+\alpha(T)P_2+\beta P_2^3+\gamma P_2^5-g_{44}\Delta P_2-f_{44}\Delta U_2=0 \quad \text{(A.6b)}$$

Boundary conditions for polarization and strain could be obtained from the variation of Eq.(A.3). In particular case $v_{ijklmn}=0$ they acquire the form:

$$\left.\left(c_{44}\frac{\partial U_2}{\partial x_3}+f_{44}\frac{\partial P_2}{\partial x_3}\right)\right|_{x_3=0}=0 \quad \text{(A.7a)}$$

$$\left.\left(\alpha_{S0}P_2-g_{44}\frac{\partial P_2}{\partial x_3}-f_{44}\frac{\partial U_2}{\partial x_3}\right)\right|_{x_3=0}=0 \quad \text{(A.7b)}$$

**B. Soft phonon's dispersion in a bulk of paraelectrics**

Let us consider a paraelectric or a ferroelectric in a paraelectric phase, when we can neglect the terms nonlinear on polarization I Eqs.(A.6):

$$\rho\frac{\partial^2 U_2}{\partial t^2}+M\frac{\partial^2 P_2}{\partial t^2}-\Delta\left(c_{44}U_2+f_{44}P_2\right)=0, \quad \text{(A.8a)}$$

$$\mu\frac{\partial^2 P_2}{\partial t^2}+M\frac{\partial^2 U_2}{\partial t^2}+\alpha(T)P_2-\Delta\left(g_{44}P_2+f_{44}U_2\right)=0. \quad \text{(A.8b)}$$

Boundary conditions are given by Eqs.(A.7). Let us look for the solution of Eqs.(A.8) in the form of a travelling wave:

$$P_2=\exp\left(i\left(kx_1-\omega t\right)\right)p\left(x_3\right), \qquad U_2=\exp\left(i\left(kx_1-\omega t\right)\right)u\left(x_3\right). \quad \text{(A.9)}$$

Substitution of these ansatzes into the equations (A.8) leads to the following ordinary differential equations:

$$-\rho\omega^2 u-M\omega^2 p-c_{44}\left(-k^2+\frac{\partial^2}{\partial x_3^2}\right)u-f_{44}\left(-k^2+\frac{\partial^2}{\partial x_3^2}\right)p=0, \quad \text{(A.10a)}$$

$$-\mu\omega^2 p-M\omega^2 u+\alpha(T)p-g_{44}\left(-k^2+\frac{\partial^2}{\partial x_3^2}\right)p-f_{44}\left(-k^2+\frac{\partial^2}{\partial x_3^2}\right)u=0. \quad \text{(A.10b)}$$

Let us look for the solution of Eqs.(A.10) in the form $p(x_3)=\tilde{p}\exp(-\xi x_3)$ and $u(x_3)=\tilde{u}\exp(-\xi x_3)$, where $\xi$ is the parameter determining the inverse penetration depth of the wave. The substitution gives the following system of linear equations

$$\begin{aligned}&\left(\rho\omega^2+c_{44}\left(-k^2+\xi^2\right)\right)\tilde{u}+M\,\omega^2\tilde{p}+f_{44}\left(-k^2+\xi^2\right)\tilde{p}=0\\&\left(\mu\,\omega^2-\alpha(T)+g_{44}\left(-k^2+\xi^2\right)\right)\tilde{p}+M\,\omega^2\tilde{u}+f_{44}\left(-k^2+\xi^2\right)\tilde{u}=0\end{aligned}\tag{A.11}$$

The condition of non-trivial solution existence of the system (A.11) is

$$\left(\rho\omega^2+c_{44}\left(-k^2+\xi^2\right)\right)\left(\mu\,\omega^2-\alpha(T)+g_{44}\left(-k^2+\xi^2\right)\right)=\left(M\,\omega^2+f_{44}\left(-k^2+\xi^2\right)\right)^2\tag{A.12a}$$

Eq.(A.12a) determines possible values of parameter $\xi$ and has the following evident form:

$$\begin{aligned}&\left(c_{44}g_{44}-(f_{44})^2\right)\xi^4+\\&\left(\left(\mu\,\omega^2-\alpha(T)\right)c_{44}+\rho\omega^2g_{44}-2M\,\omega^2f_{44}-2\left(c_{44}g_{44}-(f_{44})^2\right)k^2\right)\xi^2+\\&\left(\rho\omega^2-c_{44}k^2\right)\left(\mu\,\omega^2-\alpha(T)-g_{44}k^2\right)-\left(M\,\omega^2-f_{44}k^2\right)^2=0\end{aligned}\tag{A.12b}$$

The solution of (A.12b) is

$$\xi^2=k^2+\frac{-\left(\left(\mu\,\omega^2-\alpha(T)\right)c_{44}+\rho\omega^2g_{44}-2M\,\omega^2f_{44}\right)\pm\sqrt{D}}{2\left(c_{44}g_{44}-(f_{44})^2\right)}\tag{A.12c}$$

where discriminant of Eq. (A.12c) is introduced as

$$D=\left(\begin{aligned}&\left(\left(\mu\,\omega^2-\alpha(T)\right)c_{44}+\rho\omega^2g_{44}-2M\,\omega^2f_{44}\right)^2-\\&4\left(c_{44}g_{44}-(f_{44})^2\right)\left(\rho\omega^2\left(\mu\,\omega^2-\alpha(T)\right)-\left(M\,\omega^2\right)^2\right)\end{aligned}\right)\tag{A.12d}$$

As it follows from Eq.(A.12b)

$$\left(\xi^{(1)}\xi^{(2)}\right)^2=\frac{\left(\rho\omega^2-c_{44}k^2\right)\left(\mu\,\omega^2-\alpha(T)-g_{44}k^2\right)-\left(M\,\omega^2-f_{44}k^2\right)^2}{c_{44}g_{44}-(f_{44})^2}\tag{A.12e}$$

$$\left(\xi^{(1)}\right)^2+\left(\xi^{(2)}\right)^2=-\frac{\left(\left(\mu\,\omega^2-\alpha(T)\right)c_{44}+\rho\omega^2g_{44}-2M\,\omega^2f_{44}-2\left(c_{44}g_{44}-(f_{44})^2\right)k^2\right)}{\left(c_{44}g_{44}-(f_{44})^2\right)}\tag{A.12f}$$

Since the solid occupies the semi-space $x_3\ge 0$, only the exponents vanishing or at least not increasing at $x_3\to\infty$ are present:

$$p(x_3)=p^{(1)}\exp\left(-\xi^{(1)}x_3\right)+p^{(2)}\exp\left(-\xi^{(2)}x_3\right),\tag{A.13a}$$

$$u(x_3)=u^{(1)}\exp\left(-\xi^{(1)}x_3\right)+u^{(2)}\exp\left(-\xi^{(2)}x_3\right).\tag{A.13b}$$

It is easy to get the following relation from the system (A.11):

$$u^{(i)}=-\left(\frac{M\,\omega^2+f_{44}\left(-k^2+\left(\xi^{(i)}\right)^2\right)}{\rho\omega^2+c_{44}\left(-k^2+\left(\xi^{(i)}\right)^2\right)}\right)p^{(i)}\leftrightarrow\quad u^{(i)}=-\left(\frac{\mu\,\omega^2-\alpha(T)+g_{44}\left(-k^2+\left(\xi^{(i)}\right)^2\right)}{M\,\omega^2+f_{44}\left(-k^2+\left(\xi^{(i)}\right)^2\right)}\right)p^{(i)}$$

(A.13c)

## C. Soft phonon's dispersion in near the surface of a paraelectric

Substitution to the boundary conditions (A.7) gave the following system of equations for coefficients

$$\begin{cases} c_{44}\left(u^{(1)}\xi^{(1)} + u^{(2)}\xi^{(2)}\right) + f_{44}\left(p^{(1)}\xi^{(1)} + p^{(2)}\xi^{(2)}\right) = 0, \\ \alpha_{S0}\left(p^{(1)} + p^{(2)}\right) + g_{44}\left(p^{(1)}\xi^{(1)} + p^{(2)}\xi^{(2)}\right) + f_{44}\left(u^{(1)}\xi^{(1)} + u^{(2)}\xi^{(2)}\right) = 0 \end{cases} \quad \text{(A.14a)}$$

And finally

$$\begin{cases} \xi^{(1)}\left(c_{44}u^{(1)} + f_{44}p^{(1)}\right) + \xi^{(2)}\left(c_{44}u^{(2)} + f_{44}p^{(2)}\right) = 0, \\ \alpha_{S0}p^{(1)} + \xi^{(1)}\left(f_{44}u^{(1)} + g_{44}p^{(1)}\right) + \alpha_{S0}p^{(2)} + \xi^{(2)}\left(f_{44}u^{(2)} + g_{44}p^{(2)}\right) = 0. \end{cases} \quad \text{(A.14b)}$$

Using obvious consequences of Eqs.(A.13c) one could rewrite Eqs.(A.14b) as follows

$$\begin{cases} \dfrac{\xi^{(1)}p^{(1)}}{\rho\omega^2 + c_{44}\left(-k^2 + \left(\xi^{(1)}\right)^2\right)} + \dfrac{\xi^{(2)}p^{(2)}}{\rho\omega^2 + c_{44}\left(-k^2 + \left(\xi^{(2)}\right)^2\right)} = 0, \\ \left(\alpha_{S0} + \xi^{(1)}\left(\dfrac{\omega^2 f_{44}\left(f_{44}\rho\omega^2 - M c_{44}\right)}{\left(\rho\omega^2 + c_{44}\left(-k^2 + \left(\xi^{(1)}\right)^2\right)\right)c_{44}} + g_{44} - \dfrac{\left(f_{44}\right)^2}{c_{44}}\right)\right)p^{(1)} + \\ + \left(\alpha_{S0} + \xi^{(2)}\left(\dfrac{\omega^2 f_{44}\left(f_{44}\rho\omega^2 - M c_{44}\right)}{\left(\rho\omega^2 + c_{44}\left(-k^2 + \left(\xi^{(2)}\right)^2\right)\right)c_{44}} + g_{44} - \dfrac{\left(f_{44}\right)^2}{c_{44}}\right)\right)p^{(2)} = 0 \end{cases} \quad \text{(A.14c)}$$

The condition of zero determinant of the system (A.14c) gives the following equation

$$\begin{aligned} &\frac{\xi^{(1)}}{\rho\omega^2 + c_{44}\left(-k^2 + \left(\xi^{(1)}\right)^2\right)}\left(\alpha_{S0} + \xi^{(2)}\left(\frac{\omega^2 f_{44}\left(f_{44}\rho\omega^2 - M c_{44}\right)}{\left(\rho\omega^2 + c_{44}\left(-k^2 + \left(\xi^{(2)}\right)^2\right)\right)c_{44}} + g_{44} - \frac{\left(f_{44}\right)^2}{c_{44}}\right)\right) - \\ &\frac{\xi^{(2)}}{\rho\omega^2 + c_{44}\left(-k^2 + \left(\xi^{(2)}\right)^2\right)}\left(\alpha_{S0} + \xi^{(1)}\left(\frac{\omega^2 f_{44}\left(f_{44}\rho\omega^2 - M c_{44}\right)}{\left(\rho\omega^2 + c_{44}\left(-k^2 + \left(\xi^{(1)}\right)^2\right)\right)c_{44}} + g_{44} - \frac{\left(f_{44}\right)^2}{c_{44}}\right)\right) = 0 \end{aligned} \quad \text{(A.15a)}$$

After elementary transformations

$$\alpha_{S0}\left(\xi^{(1)} - \xi^{(2)}\right)\left(\rho\omega^2 - c_{44}k^2 - c_{44}\xi^{(1)}\xi^{(2)}\right) = \left(\left(\xi^{(1)}\right)^2 - \left(\xi^{(2)}\right)^2\right)\xi^{(2)}\xi^{(1)}\left(c_{44}g_{44} - \left(f_{44}\right)^2\right) \quad \text{(A.15b)}$$

It is seen that Eq.(A.15b) is equivalent to the following independent equations:

$$\alpha_{S0}\left(\rho\omega^2 - c_{44}k^2 - c_{44}\xi^{(1)}\xi^{(2)}\right) = \left(\xi^{(1)} + \xi^{(2)}\right)\xi^{(2)}\xi^{(1)}\left(c_{44}g_{44} - \left(f_{44}\right)^2\right), \quad \text{(A.16a)}$$

$$\xi^{(1)} = \xi^{(2)} \quad \text{(A.16b)}$$

**The solution of Eq.(A.16a) could be simplified** for two limiting cases, namely:

1) For the case of "natural boundary condition" for polarization (i.e. $\alpha_{S0} = 0$),

$$\left(\xi^{(1)} + \xi^{(2)}\right)\xi^{(2)}\xi^{(1)}\left(c_{44}g_{44} - \left(f_{44}\right)^2\right) = 0 \ \text{ or } \ \xi^{(2)}\xi^{(1)} = 0 \quad \text{(A.17a)}$$

Which is equivalent to

$$\left(\rho\omega^2 - c_{44}k^2\right)\left(\mu\,\omega^2 - \alpha(T) - g_{44}k^2\right) = \left(M\,\omega^2 - f_{44}k^2\right)^2 \quad \text{(A.17b)}$$

Eq. (A.17b) represents dispersion relation for bulk shear wave coupled with the transversal optical phonon mode.

For the case of $\alpha_{S0} = \infty$ (zero polarization at the surface and so $p^{(1)} = -p^{(2)}$)

$$\rho\omega^2 - c_{44}k^2 - c_{44}\xi^{(1)}\xi^{(2)} = 0 \tag{A.18a}$$

Using relation (A.12e) for $\left(\xi^{(1)}\xi^{(2)}\right)^2$, one could rewrite Eq.(A.18a) as

$$\frac{\left(\rho\omega^2 - c_{44}k^2\right)^2}{c_{44}^2} = \frac{\left(\rho\omega^2 - c_{44}k^2\right)\left(\mu\omega^2 - \alpha(T) - g_{44}k^2\right) - \left(M\omega^2 - f_{44}k^2\right)^2}{c_{44}g_{44} - \left(f_{44}\right)^2} \tag{A.18b}$$

**The evident form of Eq.(A.16b),** which is equivalent to the condition of zero discriminant (A.12d), is

$$\left(\left(\mu\omega^2 - \alpha(T)\right)c_{44} + \rho\omega^2 g_{44} - 2M\omega^2 f_{44}\right)^2 = 4\left(c_{44}g_{44} - \left(f_{44}\right)^2\right)\left(\rho\omega^2\left(\mu\omega^2 - \alpha(T)\right) - \left(M\omega^2\right)^2\right) \tag{A.19}$$

Note, that the solution of Eq.(A.19) with respect to frequency is independent on the wave vector (similar solution was found by Romeo [17]). It means that "frequency dispersion" is limited to the discrete set of frequency values rather than the functional dependence $\omega(k)$. This secular case is considered in details below.

**D. The secular solution in case of $\xi^{(1)} = \xi^{(2)} \equiv \xi$ under the condition (A.19)**

Here one should change the general form of the solution (A.13) as follows:

$$p(x_3) = \left(p^{(1)} - p^{(2)}\xi x_3\right)\exp(-\xi x_3) \text{ and } u(x_3) = \left(u^{(1)} - u^{(2)}\xi x_3\right)\exp(-\xi x_3) \tag{A.20}$$

with "decreasing" exponents only and new designation as it follows from (A.12c)

$$\xi^2 = k^2 - \frac{\left(\mu\omega^2 - \alpha(T)\right)c_{44} + \rho\omega^2 g_{44} - 2M\omega^2 f_{44}}{2\left(c_{44}g_{44} - \left(f_{44}\right)^2\right)}. \tag{A.21}$$

Substitution of (A.20) to the boundary conditions (A.7) gave the following system of equations for unknown coefficients

$$\begin{aligned} &c_{44}\left(u^{(1)} + u^{(2)}\right) + f_{44}\left(p^{(1)} + p^{(2)}\right) = 0 \\ &\alpha_{S0}p^{(1)} + \xi\left(g_{44}\left(p^{(1)} + p^{(2)}\right) + f_{44}\left(u^{(1)} + u^{(2)}\right)\right) = 0 \end{aligned} \tag{A.22}$$

One could easily show that relations (A.13c) are still valid for the solution (A.21). Therefore, (A.22) could written as:

$$\begin{aligned} &\left(\frac{-c_{44}M\omega^2 + f_{44}\rho\omega^2}{\rho\omega^2 + c_{44}\left(-k^2 + \xi^2\right)}\right)\left(p^{(1)} + p^{(2)}\right) = 0 \\ &\alpha_{S0}p^{(1)} + \xi\left(\frac{\omega^2 f_{44}\left(f_{44}\rho\omega^2 - Mc_{44}\right)}{\left(\rho\omega^2 + c_{44}\left(-k^2 + \xi^2\right)\right)c_{44}} + g_{44} - \frac{\left(f_{44}\right)^2}{c_{44}}\right)\left(p^{(1)} + p^{(2)}\right) = 0 \end{aligned} \tag{A.23}$$

The condition of zero determinant of the system (A.23) is

$$\left(\frac{-c_{44}M\,\omega^2 + f_{44}\rho\omega^2}{\rho\omega^2 + c_{44}\left(-k^2 + (\xi)^2\right)}\right)\alpha_{S0} = 0 \tag{A.24}$$

It is the condition of the surface wave existence in the secular case of $\xi^{(1)} = \xi^{(2)} \equiv \xi$ (under the condition (A.19)). It is seen that (A.24) could be reduced to the very specific conditions

$$\left(-c_{44}M + f_{44}\rho\right)\omega^2 = 0 \tag{A.25a}$$

$$\alpha_{S0} = 0 \tag{A.25b}$$

Thus a detailed consideration of the case leads to the conclusion that the surface wave can exist under the validity of very specific equalities (A.25), $\left(f_{44}\rho - c_{44}M\right)\omega^2 = 0$ or $\alpha_{S0} = 0$. Since there is no grounds to suppose that $f_{44}\rho - c_{44}M = 0$, the condition $\omega^2 = 0$ should be valid. However $\omega^2 = 0$ is not compatible with the condition (13), because $B(\omega = 0) = -\alpha(T)c_{44} \neq 0$ for the case. Hence we should assume $\alpha_{S0} = 0$ for the wave existence. Following Eqs.(10) $\xi^2 = k^2 - \frac{B(\omega)}{2\left(c_{44}g_{44} - (f_{44})^2\right)}$, and so the inequality $k^2 \geq \frac{B(\omega)}{2\left(c_{44}g_{44} - (f_{44})^2\right)}$ should be valid for the fulfillment of the physical condition $\mathrm{Re}(\xi) \geq 0$.